# *Ab initio* computation of Auger decay in heavy metals: zinc about it.

*Anthuan Ferino-Pérez*\*, *Thomas-C. Jagau*

**ABSTRACT**

We report the first coupled-cluster study of Auger decay in heavy metals. The zinc atom is used as a case study due to its relevance to the Auger emission properties of the $^{67}$Ga radionuclide. Coupled-cluster theory combined with complex basis functions is used to describe the transient nature of the core-ionized zinc atom. We also introduce second-order Møller-Plesset perturbation theory as an alternative method for computing partial Auger decay widths. Scalar-relativistic effects are included in our approach for computing Auger electron energies by means of the spin-free exact two-component one-electron Hamiltonian, while spin-orbit coupling is treated by means of perturbation theory. We center our attention on the K-edge Auger decay of zinc dividing the spectrum into three parts (K-LL, K-LM, and K-MM) according to the shells involved in the decay. The computed Auger spectra are in good agreement with experimental results. The most intense peak is found at an Auger electron energy of 7432 eV, which corresponds to a $^1D_2$ final state arising from K-$L_2L_3$ transitions. Our results highlight the importance of relativistic effects for describing Auger decay in heavier nuclei. Furthermore, the effect of a first solvation shell is studied by modeling of Auger decay in the hexaaqua-zinc (II) complex. We find that K-edge Auger decay is slightly enhanced by the presence of the water molecules as compared to the bare atom.

# INTRODUCTION

The Auger-Meitner effect[1-2] is a non-radiative phenomenon in which a core-ionized or a core-excited species decays through a two-electron mechanism into a doubly or singly ionized state, respectively. One electron fills the hole in the core orbital while another electron, called the Auger electron, is emitted into the continuum. Core-vacant states can be created by X-ray irradiation of molecules composed of light atoms[3] or by nuclear processes such as electron capture and internal conversion in proton-rich radionuclides.[4] If a core-ionized state is formed, we call the ensuing process non-resonant Auger decay,[3] while the corresponding process that core-excited states undergo is called resonant Auger decay.[5-6] Less common are double and triple Auger decay where 2 or 3 electrons are simultaneously emitted.[7-8] Since a core orbital is involved in the process, Auger decay is sensitive to the nature of the studied system which makes Auger spectroscopy[3] a useful tool for the thorough characterization of surfaces,[9-10] materials,[11-12] nanostructures,[13-15] and molecules.[16-19]

Furthermore, the emitted Auger electron usually triggers multiple electron tracks in the vicinity (typically 2-500 nm) of the emission site.[3] The short range of these phenomena results in high levels of energy transfer to the surrounding molecules, which allows Auger emitters to deliver considerable levels of radiation to a specific target.[4, 20-21] **Figure 1** illustrates the difference in the amount of radiation delivered to a biological target by different types of ionizing radiation. In comparison to other types of radiation employed in radiotherapy (i.e., α and β⁻ rays), Auger emitters coupled with appropriate molecular carriers are the ideal candidates for delivering high levels of radiation to tumor-specific biomarkers while avoiding off-site damage to healthy tissues. In the last decade, radiopharmaceuticals labeled with $^{111}$In and $^{125}$I have reached the stage of clinical trials,[22-23] which has placed Auger emitters as a promising alternative for radiotherapy and precision medicine.[24-25]

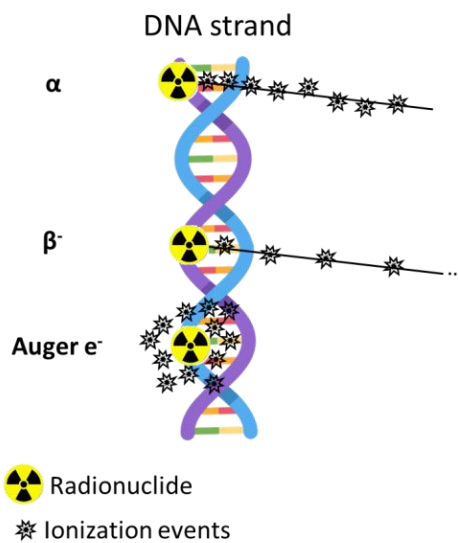

**Figure 1.** Schematic representation of the amount of energy delivered to a biological target by different types of ionizing radiation used in radiotherapy. A DNA double strand is taken as an example of a biological target.

The most common Auger-emitting isotopes available for such applications are $^{123}$I, $^{125}$I, $^{67}$Ga, $^{99m}$Tc, $^{111}$In, and $^{201}$Th. Among them, $^{67}$Ga has several benefits when compared to other candidates. Notably, there is the possibility of forming a theragnostic pair with the generator-produced positron emitter $^{68}$Ga or with itself as $^{67}$Ga has been used in single-photon emission computerized tomography (SPECT) radiodiagnosis for more than 50 years. Moreover, advances in gallium chelator synthesis, motivated by the interest in increasing the specific activity of $^{68}$Ga formulations used in positron emission tomography (PET), have enabled the efficient production of therapy-oriented radiopharmaceuticals labeled with $^{67}$Ga. Also, Auger electrons emitted after $^{67}$Ga decay are among the most energetic ones and have a relatively long electron range, which reduces the necessity that the radionuclide reaches a specific subcellular compartment to have the desired therapeutic effect.[26] The emission of the Auger electron takes place after a core vacancy is created in the 1s orbital (K shell) by electron capture:

$$^{67}_{31}\text{Ga} + e^- \xrightarrow{EC} {}^{67}_{30}\text{Zn} + \nu_e \tag{1}$$

where $\nu_e$ denotes a neutrino. This nuclear reaction produces the actual emitter of the Auger electron, $^{67}$Zn. For that reason, our attention in this work will center around K-edge Auger decay of zinc. We note, however, that vacancies in the L shell can decay non-radiatively as well.

The core-vacant states that undergo Auger decay are metastable since they have a finite lifetime directly related to this electronic decay process. Hence, they are embedded in the continuum and should be described as electronic resonances.[27-28] Therefore, the standard theoretical methods designed to tackle bound electronic states are not suitable for studying Auger decay. Furthermore, in nuclei as heavy as zinc, theoretical modeling of Auger decay requires consideration of relativistic effects. Although modern relativistic quantum chemistry offers a variety of techniques for accurate modeling,[29] little work has been done about the treatment of metastable states and Auger decay in a relativistic context. Also, these previous investigations remained limited to atoms.

K-shell Auger decay rates of various atoms with $18 < Z < 96$ have been computed using Dirac-Hartree-Slater (DHS) theory[30] as well as configuration interaction.[31] Later on, Dirac-Hartree-Fock (DHF) theory was used to compute K-shell Auger decay rates and $L_1$-shell Coster-Kronig decay rates of atoms with $32 < Z < 92$.[32] More recently, multiconfiguration DHF theory was employed to compute K-shell Auger decay rates of zinc[33] and copper[34] atoms. All these investigations used golden-rule expressions for the decay rate that rely on perturbation theory and a partitioning of the Hilbert space into a bound part and a continuum part.[35-36]

Recently, a new theoretical framework[37-38] was introduced to compute Auger decay rates using complex-scaled electronic-structure methods and was subsequently used to model Auger spectra of ethane, ethylene, and acetylene[37-40] as well as benzene.[37-40] Complex scaling[41-42] affords an $L^2$ representation of the resonance wavefunction through analytical continuation of the Hamiltonian to the complex plane. In the case of molecular resonances described in Gaussian basis sets, this is preferably done by means of the method of complex basis functions (CBFs).[43-45] Here, complex scaling is not applied to the Hamiltonian, but to the exponents of selected basis functions. In contrast to complex scaling of the Hamiltonian, the CBF method is compatible with the Born–Oppenheimer approximation, which enables the combination with modern bound-state electronic-structure methods.[28] In a basis set that includes complex-scaled functions, the eigenvalues of the Hamiltonian are no longer exclusively real, but can take complex values as well. These complex eigenvalues

$$E_{res} = E_R - i\Gamma/2 \qquad (2)$$

can be associated with resonances. The real part ($E_R$) describes the energy of the resonance state, while the imaginary part ($\Gamma$) describes its width, i.e., the rate of decay. An advantage of the CBF method is that $\Gamma$ is obtained without making any assumption about the wave function of the outgoing electron, which makes it a true *ab initio* approach.

In the present work, we extend the modeling of Auger spectra based on complex-scaled wavefunctions to the K-edge of zinc. The comparison with experimental results allows us to test if the theoretical protocol[37-40] that we developed for light nuclei is also valid for heavier nuclei that are relevant to radiomedicine. We use non-relativistic coupled-cluster singles and doubles (CCSD) theory with CBFs to compute the Auger decay rates and equation-of-motion (EOM) CCSD theory combined with the spin-free exact two-component 1-electron (SFX2C-1e) approach to compute the Auger electron energies. These first-principles post-Hartree-Fock calculations are insightful not only on their own but also help validate simpler approaches for describing Auger decay. In addition, we investigate the influence of an aqueous solvation shell on the decay rates to examine to what degree the chemical environment modulates the Auger spectrum. These calculations also serve to illustrate that our approach can be easily extended from atoms to molecules. The structure of the remainder of the article is as follows: Section 2 presents the relevant theoretical and computational details, while in Section 3 the Auger spectrum of zinc is discussed and our results are compared with other available experimental and theoretical data. Section 4 offers our concluding remarks.

**THEORETICAL AND COMPUTATIONAL DETAILS**

**Electronic structure of the zinc atom and the hexaaqua-zinc (II) complex.** Zinc is the last of the first-row transition metals. Its electron configuration is $[Ar]3d^{10}4s^2$, the term symbol of the ground state is $^1S_0$. Spin-orbit interaction is significant for the zinc atom and leads to a measurable splitting of most of its electronic states. **Figure 2** depicts the electronic structure including spin-orbit splitting and, additionally, establishes the connection between the atomic notation and the X-ray notation commonly used in core spectroscopy of heavier elements.

We separate the one-electron levels of the zinc atom into three main groups: core orbitals, inner valence orbitals, and outer valence orbitals. We only consider the 1s orbital (K shell) as core orbital. The 2s and 2p levels that belong to the L shell form the inner valence orbitals while the outer valence orbitals comprise the 3s, 3p, and 3d orbitals (M shell) and the 4s orbital (N shell). In the following, we will use the X-ray notation to name shells and transitions. For a given transition K-$X_nY_m$, K represents the core hole, while $X_nY_m$ denotes the orbitals that are initially occupied by the electrons involved in the decay. Specific references to orbitals using the atomic notation will be used for clarity when needed.

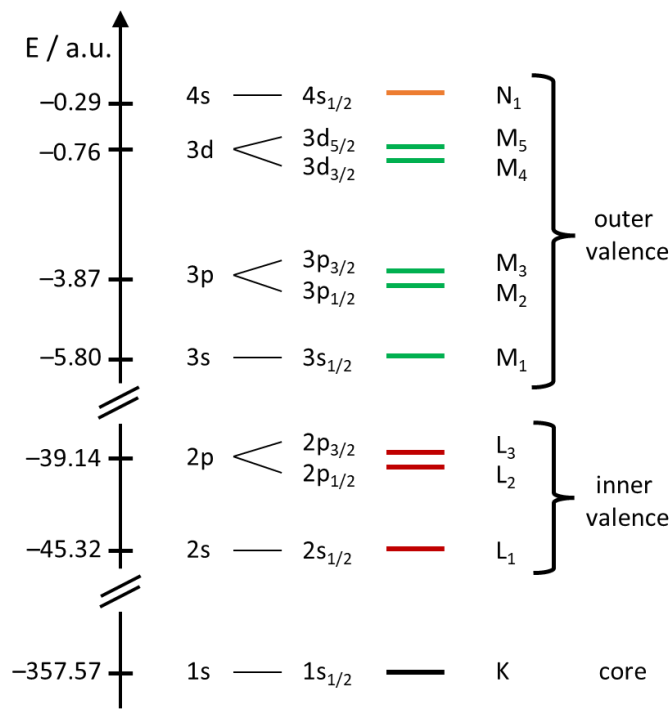

**Figure 2.** Electronic structure of the zinc atom. The labeling on the left follows the atomic notation, while the labeling on the right follows the X-ray notation. Orbital energies were computed with HF theory using the ANO-RCC basis set.[46]

It is possible to identify three distinct regions in the K-edge Auger spectrum of zinc. The slowest Auger electrons are produced by K-LL transitions, which only involve inner valence orbitals. The second region of the spectrum includes both K-LM and K-LN transitions, accounting for decay channels where one inner valence and one outer valence orbital are involved. The highest energy Auger electrons are produced by transitions involving no inner valence shells, these include the K-MM, K-MN, and K-NN transitions. All these regions need to be considered to achieve a complete picture of the K-shell Auger decay. Using the orbital energies from **Figure 2**, it is possible to get an estimate for the kinetic energy of the Auger electrons, that is, the positions of the peaks in the Auger spectrum, from

$$E_{Auger} = E_{v1} + E_{v2} - E_{core} \qquad (3)$$

where $E_{v1}$ and $E_{v2}$ are the energies of the valence levels involved in the decay and $E_{core}$ is the energy of the core orbital. Eq. (3), which neglects relativistic effects, orbital relaxation, and electron correlation, suggests energy ranges of 7260-7600 eV for the first region of the spectrum (K-LL), of 8340-8660 eV for the second region of the spectrum (K-LM and K-LN), and of 9410-9710 eV for the third region of the spectrum (K-MM, K-MN, and K-NN). As our results in Section 3 illustrate, this is qualitatively correct, but at the same time, many energies are off by more than 100 eV.

In aqueous solution, the $Zn^{2+}$ cation forms, among other species, hexacoordinated complexes with the water molecules of the environment.[47-48] We investigated the hexaaqua-zinc (II) complex ($[Zn(H_2O)_6]^{2+}$) to study the effect of a solvation shell on Auger decay. The electronic structure of this metal aquo complex is very similar to that of the bare zinc atom. The presence of the ligands leads to a symmetry lowering from SO(3) to the $T_h$ point group, but the orbital energies do not change significantly. The geometry of $[Zn(H_2O)_6]^{2+}$, optimized using RI-MP2/cc-pVTZ[49], is reported in the Supporting Information. Our calculations on Zn and $[Zn(H_2O)_6]^{2+}$ were run in the $D_{2h}$ point group, which is the largest Abelian subgroup of the SO(3) and $T_h$ point groups.

**Positions of the peaks in the Auger spectrum.** For computing a theoretical Auger spectrum from first principles, it is necessary to determine both the position and the intensity of the peaks, which is done in separate calculations in our approach.[39] **Figure 3** illustrates the states that need to be considered.

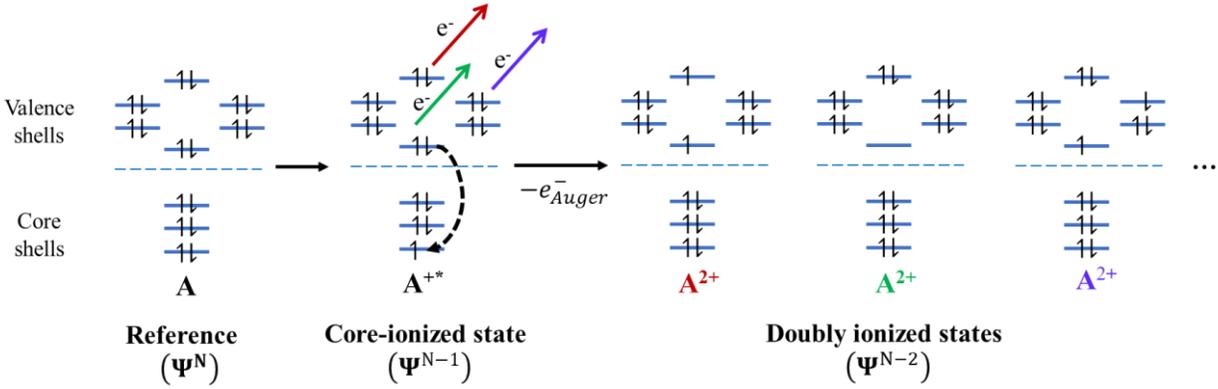

**Figure 3.** Electronic states relevant to *ab initio* modelling of Auger spectra. N denotes the number of electrons in the system.

The position of a given peak in the Auger spectrum corresponds to the kinetic energy of the Auger electron, which can be calculated from the energies of the core-ionized (IP) and double-ionized (DIP) states as

$$E_{Auger} = E_{DIP} - E_{core\ IP} \qquad (4)$$

Eq. (4) reduces to Eq. (3) in the framework of HF theory or density functional theory but represents an improvement in the context of correlated calculations.

In this work, the energies of the core-ionized and doubly ionized states were obtained from EOM-CCSD calculations[50-54] in the ionization potential[51, 53] (EOM-IP-CCSD) and double ionization potential[55-56] (EOM-DIP-CCSD) variants, respectively. The ground state of the zinc atom was used as CCSD reference state in all these calculations. To take account of scalar-relativistic effects, we use the spin-free exact two-component one-electron (SFX2C-1e) scheme.[57] In this single-component method, scalar relativity is captured by using modified one-electron and two-electron integrals, which means that non-relativistic electronic-structure codes can be used after modifications at the level of the integral evaluation. Spin-orbit coupling (SOC) is treated in our calculations *a posteriori* as a perturbation to the SFX2C-1e/EOM-CCSD Hamiltonian.[58-59]

There is, however, no implementation of SFX2C-1e/EOM-DIP-CCSD available. Therefore, we replaced in this work the usual EOM-DIP-CCSD excitation operator:

$$R_{DIP} = \frac{1}{2}\sum_{ij} r_{ij}\{ij\} + \frac{1}{6}\sum_{a}\sum_{ijk} r_{ijk}^{a}\{a^{\dagger}ijk\} \tag{5}$$

by a truncated version without 3-hole 1-particle excitations. An EOM-CCSD calculation with the operator

$$R_{DIP}^{mod} = \frac{1}{2}\sum_{ij} r_{ij}\{ij\} \tag{6}$$

can be carried out using an EOM-EE-CCSD code and the continuum orbital trick,[60] but the resulting wave functions are essentially uncorrelated, and the corresponding energies are therefore expected to be substantially too high. These EOM-CCSD/SFX2C-1e calculations were carried out in the ANO-RCC basis[46] using the CFOUR program package.[61]

In addition, we carried out non-relativistic EOM-DIP-CCSD calculations based on Eq. (5) with the Q-Chem 6.0 software package.[62] The values for $E_{DIP}$ obtained in these latter calculations are off by up to 100 eV, which shows that the inclusion of scalar relativity is vastly more important than a proper treatment of electron correlation. In fact, our results presented in Section 3 suggest that the switch from Eq. (5) to Eq. (6) largely comes down to a global shift of all peak positions by a few eV. We also note that the non-relativistic EOM-DIP-CCSD states obtained using Eq. (5) are subject to substantial configuration mixing, which led to convergence problems of the Davidson algorithm in many instances. The calculations with the reduced excitation manifold from Eq. (6), on the other hand, do not suffer from any convergence problems.

**Intensity of the peaks in the Auger spectrum.** The intensity of a peak in the Auger spectrum is related to the width of the corresponding decay channel. In this work, we determine these partial decay widths by analyzing the imaginary part of the CBF-CCSD energy of the core-ionized state.[37]

A new basis set denoted aug-cc-pCVTZ(5sp)+4(spd)2f, whose details are discussed in the subsequent section, was constructed for this purpose. All CBF-CCSD calculations were done without considering relativistic effects using the complex-variable CCSD codes[45, 63-65] in the Q-Chem 6.0 software package.[62]

The complex energies from Eq. (2) obtained from CBF calculations in a finite basis set are subject to an unphysical dependence on the complex-scaling angle θ. In the present case, the optimal value of θ was determined to be 19° by minimizing $|d(E_{CCSD}^{N-1} - E_{CCSD}^{N})/d\theta|$,[66] where $E_{CCSD}^{N-1}$ and $E_{CCSD}^{N}$ are the CBF-CCSD energies of the core-ionized state and the reference state, respectively. At this complex-scaling angle, the total decay width was computed according to

$$\Gamma = -2 \cdot \left[\text{Im}(E_{CCSD}^{N-1}) - \text{Im}(E_{CCSD}^{N})\right] \quad (7)$$

We note that in the limit of a full basis set, $\text{Im}(E_{CCSD}^{N})$ would vanish, since the ground state of zinc is not subject to decay. However, taking $\text{Im}(E_{CCSD}^{N})$ into account in Eq. (7) removes errors introduced by the finite basis set and thus leads to a more balanced description.[37, 65, 67]

The partial decay width $\gamma_{ij}$ of a channel involving occupied orbitals i and j was calculated as[37]

$$\gamma_{ij} = -2 \cdot \text{Im}\left(\sum_{a}^{\text{virt}} \left(\frac{1}{2}t_{ij}^{ab} + t_i^a t_j^b\right) \langle ij||ab\rangle\right) \quad (8)$$

Note that Eq. (8) does not involve the wave function nor the energy of the doubly ionized final state. Rather, all information necessary to determine the partial decay width is contained in the complex energy of the core-ionized state.

Eqs. (7) and (8) apply not only to CBF-CCSD calculations, but to CBF-MP2 calculations as well. In the latter case, Eq. (8) reduces to

$$\gamma_{ij} = -\text{Im}\left(\sum_a^{\text{virt}} \frac{[\langle ij|ab\rangle - \langle ij|ba\rangle]^2}{\varepsilon_a + \varepsilon_b - \varepsilon_i - \varepsilon_j}\right) \qquad (9)$$

In the present work, we use the CBF-MP2 method for computing the decay widths of the $[Zn(H_2O)_6]^{2+}$ complex. Notably, the optimal complex scaling angle was still 19° in these calculations, i.e., the presence of the water molecules and the switch from CBF-CCSD to CBF-MP2 do not lead to a significant change.

To construct the Auger spectrum, the partial decay widths obtained from Eqs. (8) and (9) were linked to the double ionization energies by assigning to a given EOM-DIP-CCSD state an intensity equal to the weighted sum of those $\gamma_{ij}$ values for which i and j correspond to the leading EOM-DIP-CCSD $r_{ij}$ amplitudes. The amplitudes $r_{ij}$ were also used as weighting coefficients in this procedure.[39] The final spectra were then built by applying a Lorentzian broadening function to all decay channels. A $\sigma$ value of 7 eV was chosen for all decay channels to match the experimental resolution.

**Basis set details.** The computation of total and partial widths with the CBF-CCSD method requires basis sets that correctly describes core correlation and, at the same time, the outgoing Auger electron. In recent work involving only elements from the second row of the periodic table,[37-40] we constructed such bases starting from Dunning's correlation consistent basis sets cc-pCV*X*Z and aug-cc-pCV*X*Z.[68-70] We found that replacing the *s* and *p* shells of the aug-cc-pCVTZ basis set by those from the aug-cc-pCV5Z basis set improves the description of Auger decay.[37-40] Because neither aug-cc-pCVTZ nor aug-cc-pCV5Z are available for zinc, we constructed an aug-cc-pCVTZ(5sp) basis set starting from aug-cc-pwCVTZ.[71] This latter basis set accounts for some core correlation, but its exponents were optimized weighting valence-core correlation over core-core correlation, hence wC (weighted core) in its name. The procedure from Refs. 70-72, which minimizes the core-correlation energy

$$E_{corr}^{core} = E_{corr} - E_{corr}^{frozen\ core} \qquad (9)$$

was used to remove the bias in the basis set and account for core-core correlation as well.

To the aug-cc-pCVTZ(5sp) basis set, we added four complex-scaled $s$, $p$, and $d$ shells and two complex-scaled $f$ shells, whose exponents were optimized by minimizing $|d(E_{CCSD}^{N-1} - E_{CCSD}^{N})/d\theta|$.[37] The resulting basis set, denoted aug-cc-pCVTZ(5sp)+4(spd)2f, was used for all complex-variable calculations on the zinc atom and is reported in the Supporting Information. In the case of the hexaaqua-zinc (II) complex, we used the aug-cc-pCVTZ(5sp)+4(spd)2f basis set for the zinc atom and similar aug-cc-pCVTZ(5sp)+2(spd) basis sets for the oxygen and hydrogen atoms. The latter were taken from Ref. 38 and not re-optimized in this work. In total, 1331 basis functions were included in the CBF-MP2 calculations on $[Zn(H_2O)_6]^{2+}$.

Note that for decay channels involving the water molecules, the Auger electron can only originate from molecular orbitals formed by 1s, 2s, or 2p atomic orbitals. Hence, the angular momentum of the Auger electron is dominated by the quantum numbers $l$=0-2 (*s-d*). The zinc atom, on the other hand, also has occupied 3d orbitals giving rise to Auger electrons of higher angular momentum quantum numbers up to, in principle, *l=4 (g)*. We found, however, that already the complex-scaled f-shells have limited impact, which is why we decided against the inclusion of complex-scaled g-shells in the basis set.

## RESULTS AND DISCUSSION

**Core-ionization energy.** The computation of the core-ionization energy (IE) is a natural test of the quality of the description of the core-ionized state. We computed two values for the zinc atom: The first calculation used non-relativistic EOM-IP-CCSD, while the second one combined EOM-IP-CCSD with the SFX2C-1e Hamiltonian that accounts for scalar-relativistic effects. The results are shown in **Table 1**. The high values illustrate how much energy is needed to remove an electron from a core orbital in a heavy element like zinc as compared to first-row elements like carbon (IE = 290.84)[73] or oxygen (IE = 543.28 eV).[73] This is one of the main reasons why nuclear processes such as electron capture that takes place in proton-rich nuclei are more efficient for the formation of Auger emitters in heavy elements. The core IE calculated with the non-relativistic Hamiltonian differs by almost 100 eV from the reported experimental data.[74-75] This result is improved when the SFX2C-1e Hamiltonian is used, reducing the deviation from the experimental data to 14 eV, which is similar to the difference between the two reported experimental IE values (8.4 eV). These results show the importance of scalar relativistic effects for computing accurate core IEs and, consequently, accurate peak positions in the final Auger spectrum of zinc.

**Table 1.** Core ionization energy of the zinc atom

|  | Core ionization energy (eV) |
|---|---|
| Experiment (1966)[74] | 9659.0 |
| Experiment (1977)[75] | 9667.4 |
| EOM-IP-CCSD | 9568.6 |
| EOM-IP-CCSD/SFX2C-1e | 9681.8 |

**Total decay width and branching ratios.** From the electronic structure of the zinc atom, it can be deduced that, disregarding spin degeneracy, there are a total of 196 possible Auger transitions when the initial hole is in the K shell (1 core hole × 14 valence electrons × 14 valence electrons). Following Equation (7) we obtained for the total decay width a value of 815.7 meV using CBF-CCSD and of 869.9 meV using CBF-MP2. The partial widths for all computed Auger transitions are given in Tables S1-S5 of the Supporting Information.

The Auger spectrum of zinc can be divided into three regions according to the involved orbitals and their energetic separation: K-LL, K-LM + K-LN, and K-MM + K-MN + K-NN. **Table 2** presents the cumulated decay widths of these transitions computed with CBF-CCSD and CBF-MP2 as well as a comparison with previous theoretical results.[31-33] This shows that the 16 K-LL decay channels account for about 80% of the overall intensity, while the 72 K-LM channels account for 18-20% and the remaining 108 channels (K-LN, K-MM, K-MN, K-NN) for less than 2%. The dominance of the K-LL decay channels has been ascribed to the higher spatial overlap of L electrons with the core hole.[3] Transitions involving the N shell (4s orbital) of zinc are so weak that they are not expected to have a significant impact on the Auger spectrum. This is in agreement with the reported experiments, which barely detected peaks related to these transitions.[76-77]

CBF-MP2 appears to overestimate the CBF-CCSD results systematically by about 5%. As documented in the Supporting Information, a similar pattern is observed for most individual decay channels with the mean relative error being below 10 % except for the very weak K-MN and K-NN decay channels. The biggest absolute deviation between CBF-MP2 and CBF-CCSD occur for the K-LL decay channels, which is no surprise in view of their overall dominance. The largest absolute deviation of 2.5 meV is observed for the K-$L_2L_3$ ($2p^{-1}2p^{-1}$) channels. In general, it can be concluded that CBF-MP2 is a valid alternative to CBF-CCSD for the computation of K-edge Auger decay widths.

We note that the sum of the CBF-CCSD and CBF-MP2 partial widths in **Table 2** is not identical to the total widths computed according to Equation (7). This is due to the contributions of unphysical decay channels, which do not describe Auger decay, to the total decay width as well as the imaginary part of the energy of the neutral zinc atom.[37]

**Table 2.** Auger decay widths in meV of the different regions of the Auger spectrum of the zinc atom.

| Transitions | No. of transitions | DHS, (1980)[31] | DHF, (2001)[32] | MCDHF, (2020)[33] | CBF-CCSD, this work | CBF-MP2, this work |
|---|---|---|---|---|---|---|
| K-LL | 16 | 653.56 | 608.94 | 630.16 | 623.11 | 656.09 |
| K-LM | 72 | 163.40 | 150.15 | 164.63 | 143.17 | 149.38 |
| K-LN | 8 | 2.67 | - | 2.12 | 1.68 | 1.65 |
| K-MM | 81 | 10.99 | 9.03 | 10.99 | 7.82 | 8.26 |
| K-MN | 18 | 0.38 | - | 0.54 | 0.21 | 0.21 |
| K-NN | 1 | - | - | - | $2.05 \cdot 10^{-3}$ | $1.72 \cdot 10^{-3}$ |
| Total[a] | 196 | 829.51 | 768.20 | 807.22 | 775.94 | 815.58 |

[a]Calculated as sum of partial widths.

Our CBF-CCSD and CBF-MP2 decay widths are also in good agreement with previous DHS, DHF, and MCDHF results that take into account relativistic effects but describe electron correlation at a lower level (MCDHF) or not at all (DHS and DHF). This demonstrates two things: First, CBF-CCSD and CBF-MP2 deliver accurate total decay widths for heavier elements even if relativistic effects are disregarded. Second, the neglect of electron correlation also does not appear to impair these results.

By comparing the intensities of the different regions of the Auger spectrum, it is possible to obtain branching ratios of the total decay width. For this purpose, the most intense decay channels, i.e., those in the K-LL region of the spectrum, are used as reference. **Table 3** demonstrates excellent agreement between our results for the Auger decay branching ratios of zinc and previously reported experimental and computational values. This is a further indication that our approach captures the distribution of intensity among the different regions of the Auger spectrum correctly even though relativistic effects are not considered.

Table 3. Branching ratios for the K-edge Auger spectrum of the zinc atom

|  | (K-LM + K-LN)/K-LL | (K-MM + K-MN + K-NN)/K-LL |
|---|---|---|
| Experiment (1972)[78] | 0.23 (2) | 0.013 (4) |
| Experiment (2004)[76] | 0.24 (2) | 0.017 (2) |
| DHS (1980)[31] | 0.254 | 0.015 |
| DHF (2001)[32] | 0.247 | 0.015 |
| MCDHF (2020)[33] | 0.265 | 0.016 |
| CBF-CCSD (this work) | 0.232 | 0.013 |
| CBF-MP2 (this work) | 0.230 | 0.013 |

**Equal width approximation.** Earlier works suggested to use the density of states as a proxy for the decay rates in order to bypass the explicit calculation of the latter.[79-80] This leads to reasonable Auger spectra for molecules composed of elements from the first and second period of the periodic table. However, **Table 2** demonstrates that it is not a valid approximation in the case of the K-edge Auger spectrum of zinc because the number of transitions in a particular region is not a good indicator for estimating the relative magnitude of the decay widths. The situation is different if one considers only the K-LL region. Here, the equal width approximation describes the overall shape of the spectrum correctly, although not the relative intensity of the different peaks. However, the quality of the equal width approximation deteriorates for higher Auger electron energies as the intensity of peaks involving 3d electrons is overestimated considerably. All Auger spectra computed using the equal-width approximation are reported in the Supporting Information.

**K-LL Auger spectrum. Figure 4** compares the theoretical results obtained in this work with two sets of experimental data for the K-LL Auger spectrum of zinc.[76-77] Notably, the more recent measurement only covers Auger electron energies from ca. 7410 eV to 7490 eV,[77] whereas the older measurement reports results down to 7000 eV.[76] Also, the two measurements disagree about the position of the most intense peak by 105 eV. In **Figure 4**, we used the data from Ref. 77 as reference point and shifted the older experimental values by 105 eV to lower energies.

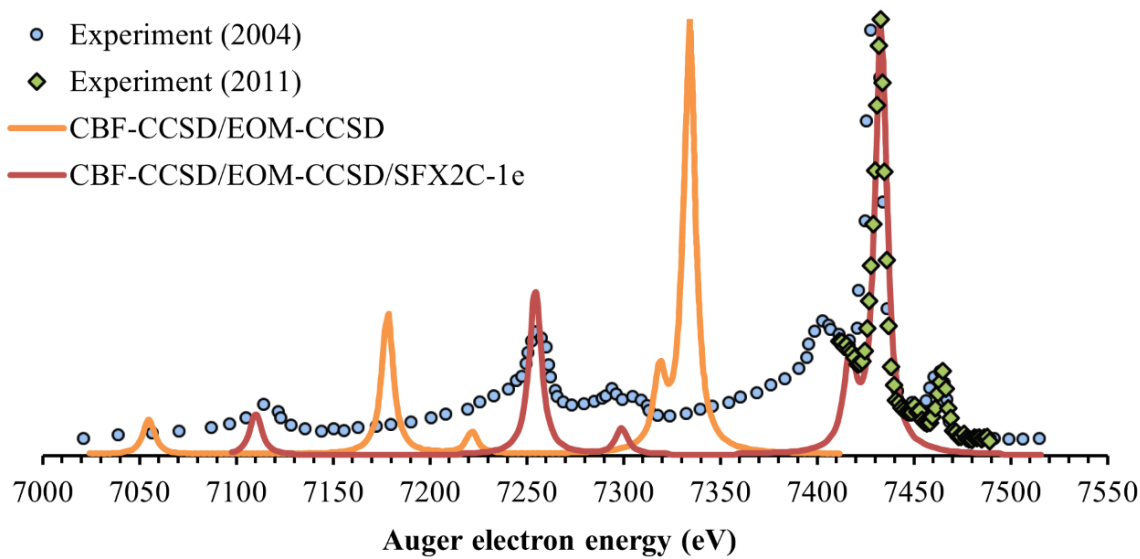

**Figure 4.** K-LL Auger spectrum of the zinc atom. Comparison between experimental data,[76-77] and CBF-CCSD results. Peak positions in the theoretical spectrum were computed using non-relativistic EOM-CCSD (orange) and SFX2C-1e/EOM-CCSD (red) with the excitation operator from Eq. (6). Both theoretical spectra are shifted by 14 eV to higher Auger electron energy, the older experimental data (blue) are shifted by 105 eV to lower Auger electron energy. All data sets are normalized to ease the comparison.

Because of the big discrepancy between the two experiments, it is difficult to ascertain the absolute accuracy of the peak positions computed with our theoretical protocol. However, **Figure 4** illustrates that inclusion of scalar relativistic effects is, in any case, indispensable. The SFX2C-1e approach increases the position of all peaks by ca. 110 eV as compared to the non-relativistic calculation. The remaining discrepancy with the experimental data[77] amounts to 14 eV; in **Figure**

**4** we accounted for it by shifting all theoretical results by 14 eV to higher energies. Notably, the core-ionization energy computed with SFX2C-1e/EOM-IP-CCSD also differs by 14 eV from the experimental value (see **Table 1**).

Despite this uncertainty about the absolute peak positions, the relative peak positions as well as the peak intensities are reproduced well by our calculations. A mismatch occurs, however, about the number of peaks: our computed spectrum comprises 5 distinct peaks, whereas 9 peaks were observed experimentally.[76]

Our theoretical K-LL Auger spectrum is built from 16 decay channels. By increasing Auger electron energy, i.e., from left to right in **Figure 4**, the first peak located at 7112 eV corresponds to the K-$L_1L_1$ ($2s^{-2}$) transition that results in a $^1S$ state of the zinc dication. The second peak at 7255 eV is formed by a K-$L_1L_2$ transition ($2s^{-1}2p^{-1}$) that yields a $^1P$ state, while the peak corresponding to the $^3P$ ($2s^{-1}2p^{-1}$) state is located at 7300 eV. The experimental spectrum shows three peaks in the latter region, likely because the $^3P$ state is subject to spin-orbit splitting. While our current approach does not allow for the consideration of SOC in the peak intensities, we can compute its effect on the peak positions. Using perturbation theory on top of our EOM-CCSD/SFX2C-1e calculations,[58-59] we obtained a value of 11.40 eV for the energy difference between the $^3P_0$ state and the $^3P_1$ state, or, equivalently, between the $^3P_1$ state and the $^3P_2$ state, which agrees well with the spacings of 10.8 eV and 12.2 eV measured in the experiment.

The last two peaks of our theoretical K-LL spectrum correspond to K-$L_2L_{2,3}$ ($2p^{-2}$) transitions. The peak at 7417 eV corresponds to a final $^1S$ state, while the one at 7432 eV corresponds to a final $^1D$ state. The latter peak is the most intense of the whole spectrum and serves as reference point in **Figure 4**. Relative to it, the position of the $^1S$ state is somewhat overestimated by our calculations, while its intensity is slightly underestimated.

The main discrepancy between theory and experiment in the K-LL spectrum is caused by the last two peaks that are measured around 7460 eV in the experiment but are not present in our theoretical spectra. A channel corresponding to decay into the $^3$P ($2p^{-2}$) state would be located at this energy, but the width of this channel is zero in our calculations. Likely, this is due to the neglect of spin-orbit coupling in the calculation of the widths, even though one would expect three peaks and not two for a $^3$P state. We computed a value of 11.42 eV for the spin-orbit splitting of the $^3$P state, which is in good agreement with the value of 13.4 eV measured in the experiment.[76] A further argument that supports the hypothesis that the two highest energy peaks are a spin-orbit effect is that the corresponding peaks are not present in Auger spectra of light elements where SOC is weaker. For example, the $^3$P ($2p^{-2}$) ground state of the neon dication, whose electronic structure is similar to that of the $^3$P state of $Zn^{2+}$, does not contribute to the Auger spectrum of neon. Also, an enhancement of the intensity of triplet decay channels by SOC has been described for Auger and Coster-Kronig decay in argon.[81-83]

**K-LM and K-LN spectrum.** The K-LM + K-LN Auger spectrum of zinc is depicted in **Figure 5**. In this case, the theoretical spectrum was shifted by 75 eV to higher Auger electron energy to match the position of the lowest energy peak in the experimental spectrum.[76] We did not use the most intense peak as reference in this case as done in **Figure 4** for the K-LL spectrum, because the experimental spectrum is dominated by two peaks of similar intensity that correspond to only one peak in our calculations. We note that the discrepancy between theory and experiment in **Figure 5** (75 eV) is actually smaller than that in **Figure 4** (119 eV). Given the discrepancy of 105 eV between the two experimental K-LL spectra,[76-77] it appears likely that a large fraction of the shift of 75 eV applied in **Figure 5** reflects inaccuracies in the measurement.

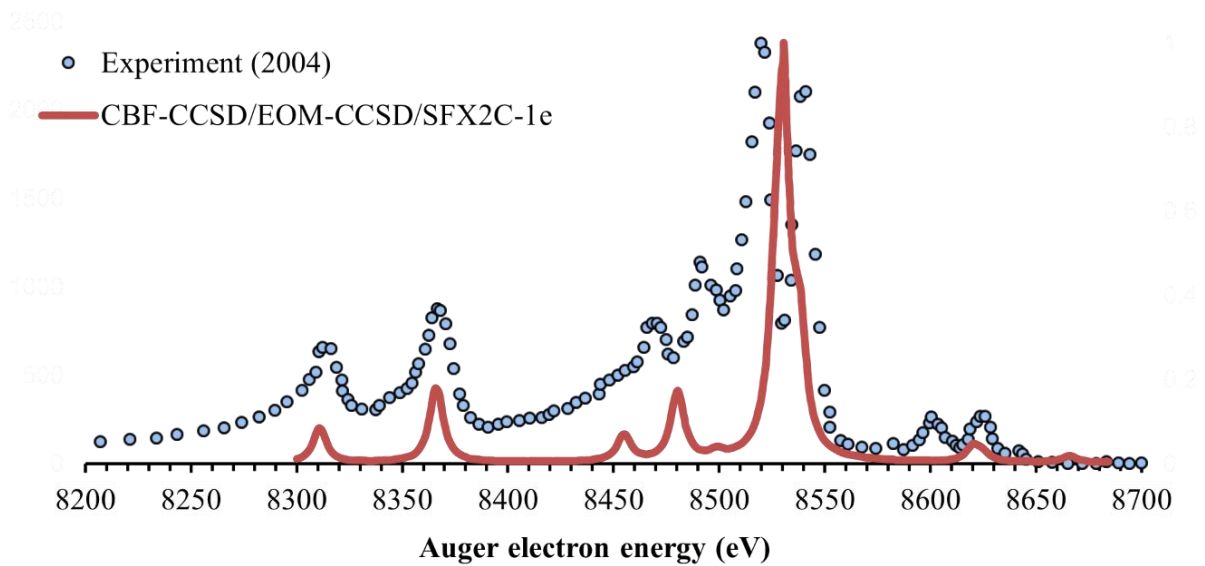

**Figure 5.** K-LM + K-LN Auger spectrum of the zinc atom. Comparison between experimental data[76] and CBF-CCSD results. Peak positions in the theoretical spectrum were computed using SFX2C-1e/EOM-CCSD with the excitation operator from Eq. (6). The theoretical spectrum is shifted by 75 eV to higher Auger electron energy. All data sets are normalized to ease the comparison.

The theoretical K-LM + K-LN Auger spectrum in **Figure 5** has 8 distinct peaks and includes 80 transitions. The intensity of most peaks is underestimated with respect to the experimental data, which is likely a consequence of the splitting of the most intense peak in the experiment as we used this peak as reference point for the normalization of the intensities. By increasing Auger electron energy, i.e., from left to right in **Figure 5**, the first three peaks account for transitions involving the 2s orbital. Specifically, the peak at 8313 eV corresponds to K-$L_1M_1$ ($2s^{-1}3s^{-1}$) transitions, the one at 8367 eV to K-$L_1M_{2,3}$ ($2s^{-1}3p^{-1}$) transitions, and the one at 8458 eV, which is only visible as a shoulder in the experimental spectrum, to K-$L_1M_{4,5}$ ($2s^{-1}3d^{-1}$) transitions. Each of these peaks includes both singlet and triplet channels as they lie very close in energy.

The fourth peak of the theoretical spectrum, located at 8491 eV, results from K-$L_{2,3}M_1$ ($2p^{-1}3s^{-1}$) transitions and includes close-lying singlet and triplet decay channels as well. In this region, the experimental spectrum has two peaks, likely again because of spin-orbit splitting. Our theoretical value for the spin-orbit splitting of the $^3P$ ($2p^{-1}3s^{-1}$) state is 11.26 eV, which is significantly smaller that the 22.6 eV extracted from the experiment. Possibly, this is because only the $^3P_0$ and $^3P_2$ components were observed in the experiment but not the $^3P_1$ component. The fifth peak of the theoretical spectrum is located at 8501 eV and consists of K-$L_1N_1$ ($2s^{-1}4s^{-1}$) channels. In Ref. 76, it was not reported, probably due to its very low intensity.

The following sixth peak of the theoretical spectrum is centered at 8530 eV and is the most intense peak of the K-LM Auger spectrum. It is composed of 15 K-$L_{2,3}M_{2,3}$ transitions, which comprise the close-lying S, P, and D singlet and triplet final states that originate from the configuration ($2p^{-1}3p^{-1}$). In the experimental spectrum, two separate intense peaks are found at 8521 and 8538 eV. According to Ref. 76, the latter of the two peaks is formed by transitions to the $^3S$, $^3P$, $^3D$, and $^1P$ states. Our calculations are not able to reproduce this splitting; the higher energy transitions only form a shoulder as their intensity is much lower than in the experiment. Notably, the $^1P$ state delivers the predominant contribution to this shoulder in our calculations, while the intensities of the triplet states are low. This suggests that the assignment of the peaks in the experiment may not be entirely correct. There is also a considerable disagreement about the spin-orbit splitting: While we obtain values of 6.36 eV and 10.39 eV for the splitting of the $^3P$ and $^3D$ states, respectively, much larger values of 20.2 eV and 26.0 eV were extracted from the experimental data.[76]

Lastly, the seventh theoretical peak at 8623 eV includes the 30 K-$L_{2,3}M_{4,5}$ ($2p^{-1}3d^{-1}$) transitions. Here again, the experimental spectrum has two peaks that are 22.0 eV apart, likely due to spin-orbit splitting. We computed for the $^3$P, $^3$D, and $^3$F states spin-orbit splittings of 5.51 eV, 9.16 eV, and 22.5 eV, respectively, so that the further analysis of the experimentally observed splitting is difficult. The last peak of the theoretical spectrum corresponds to K-L$_{2,3}$N$_1$ (2p$^{-1}$4s$^{-1}$) decay channels that are located at 8664 eV but have very small widths so that they are not visible in the measured spectrum.

**K-MM, K-MN, and K-NN spectrum. Figure 6** shows the K-MM + K-MN + K-NN Auger spectrum of zinc. The theoretical spectrum consists of 10 peaks arising from 100 transitions and was shifted by 40 eV to higher Auger electron energy to match the position of the most intense peak in the experimental spectrum.

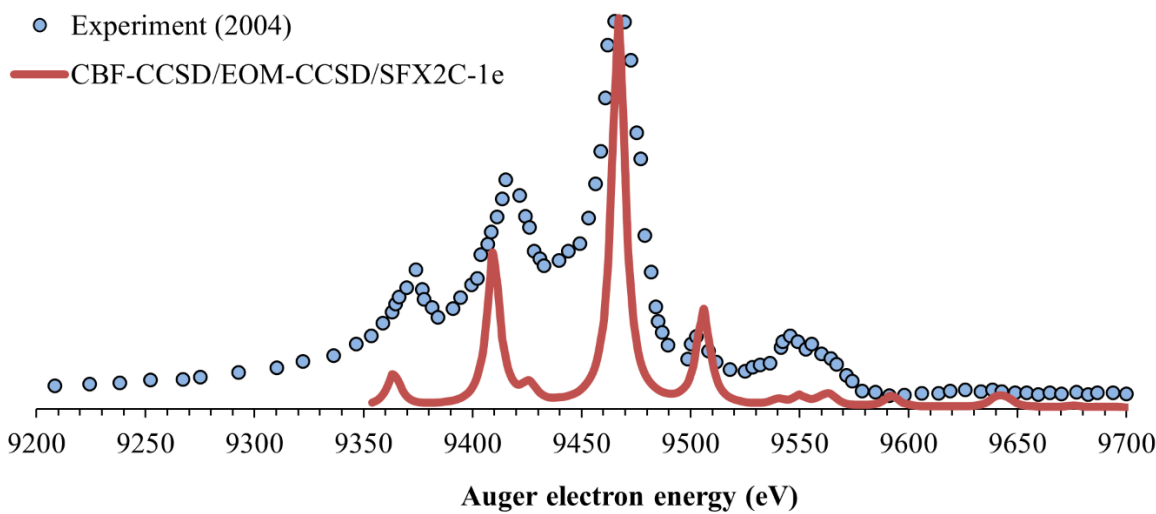

**Figure 6.** K-MM + K-MN + K-NN spectrum of the zinc atom. Comparison between experimental data[76] and CBF-CCSD results. Peak positions in the theoretical spectrum were computed using SFX2C-1e/EOM-CCSD with the excitation operator from Eq. (6). The theoretical spectrum is shifted by 40 eV to higher Auger electron energy. All data sets are normalized to ease the comparison.

The four lowest-energy peaks of the K-MM spectrum are equivalent to the K-LL spectrum and correspond to transitions involving the 3s and 3p orbitals. However, the peaks are closer together in the K-MM spectrum because the 3s and 3p orbitals are closer in energy than the 2s and 2p orbitals (see **Figure 2**). The lowest-energy peak at 9364 eV corresponds to the K-$M_1M_1$ ($3s^{-2}$) transition, whereas the next two peaks at 9408 and 9427 eV are due to K-$M_1M_{2,3}$ ($3s^{-1}3p^{-1}$) singlet and triplet decay channels. In the experimental spectrum, these transitions manifest as a single peak centered at 9415 eV. The fourth peak at 9467 eV is the most intense peak of the K-MM spectrum and results from K-$M_{2,3}M_{2,3}$ ($3p^{-1}3p^{-1}$) decay channels that correspond to $^1S$, $^3P$, and $^1D$ final states. Similar to the K-LL spectrum, the $^3P$ state has zero width in our calculations, which may change if SOC is considered.

The peak at 9506 eV is associated with K-$M_1M_{4,5}$ ($3s^{-1}3d^{-1}$) transitions, while the next three peaks, located between 9535 eV and 9570 eV, correspond to K-$M_{2,3}M_{4,5}$ ($3p^{-1}3d^{-1}$) and K-$M_1N_1$ ($3s^{-1}4s^{-1}$) transitions. None of these peaks has an analog in the K-LL spectrum. The intensity of the K-$M_1M_{4,5}$ peak is slightly overestimated compared to the experiment, whereas that of the K-$M_{2,3}M_{4,5}$ and K-$M_1N_1$ peaks is underestimated resulting in a significant discrepancy between the theoretical and experimental spectra. The last three peaks of the theoretical spectrum at 9589 eV, 9647 eV, and 9680 eV correspond to K-$M_{2,3}N_1$, K-$M_{4,5}M_{4,5}$, and K-$M_{4,5}N_1$ transitions, respectively. These decay channels have very small widths and are not visible in the experimental spectrum. Interestingly, the spin-orbit splitting appears to be relevant only for the series of peaks between 9540-9562 eV. For the rest of the spectrum, the spin-orbit splitting calculated for the triplet states is smaller than the experimental resolution (2 eV).[76]

**Hexaaqua-zinc (II) complex.** To get an estimate of the impact of a solvation shell on the Auger decay process, we studied the hexaaqua-zinc (II) complex using the CBF-MP2 method. The total decay width is 920.3 meV as compared to 869.9 meV for the bare zinc atom computed with the same approach. The presence of the solvation shell thus slightly broadens the decay width.

To analyze the effect of the solvation shell on the different decay channels, it is helpful to consider the molecular orbitals of the $[Zn(H_2O)_6]^{2+}$ complex. The corresponding plots in the Supporting Information illustrate that the 2s, 2p, 3s, and 3p orbitals of the zinc atom, which correspond to the $2a_g$, $1t_u$, $4a_g$, and $3t_u$ orbitals of $[Zn(H_2O)_6]^{2+}$, largely maintain atomic character in the complex. On the contrary, the 3d orbitals lose their atomic character and form bonding ($1t_{1g}$ + $3e_g$) and antibonding ($2t_{1g}$ + $4e_g$) molecular orbitals.

In total, we considered 1849 decay channels (1 core hole × 43 valence electrons × 43 valence electrons) of the $[Zn(H_2O)_6]^{2+}$ complex. **Table 4** compares the widths of the K-LL decay channels of the zinc atom and the hexaaqua-zinc (II) complex. These channels, all of which involve orbitals that retain atomic character in the complex, become slightly broader in the presence of the solvent shell with the largest change (11.7 meV) observed for the $2s^{-2}$ channel. A similar comparison of the decay channels that involve the 3d orbitals of the zinc atom is reported in the Supporting Information. Here, the widths slightly decrease in the presence of the solvation shell.

Decay channels involving the remaining molecular orbitals, i.e., those that cannot be mapped to the orbitals of the zinc atom but correspond to orbitals of the water molecules, together only account for a width of 1.4 meV. The water molecules thus do not participate directly in the decay process. However, for a complete picture, one needs to consider that Auger electrons emitted from the zinc atom have enough energy to core-ionize surrounding water molecules. Also, non-local decay processes such as intermolecular Coulombic decay and electron transfer-mediated decay involving water as acceptor require even less energy and are likely to occur as a follow-up process to core ionization of zinc.[84-85]

**Table 4.** Partial half-widths of the K-LL decay channels of the zinc atom and the hexaaqua-zinc (II) complex computed with CBF-MP2.

| Zn atom | | $[Zn(H_2O)_6]^{2+}$ | |
|---|---|---|---|
| Transition* (SO(3)) | Partial width [meV] | Transition* ($T_h$) | Partial widths [meV] |
| $2s^{-2}$ (S) | 41.5 | $2a_g^{-2}$ (S) | 53.2 |
| $2s^{-1}2p^{-1}$ (S) | 148.9 | $2a_g^{-1}1t_u^{-1}$ (S) | 148.9 |
| $2s^{-1}2p^{-1}$ (T) | 23.7 | $2a_g^{-1}1t_u^{-1}$ (T) | 23.7 |
| $2p^{-2}$ (S) | 193.5 | $1t_u^{-2}$ (S) | 198.4 |
| $2p^{-1}2p^{-1}$ (S) | 248.4 | $1t_u^{-1}1t_u^{-1}$ (S) | 248.8 |
| $2p^{-1}2p^{-1}$ (T) | 0 | $1t_u^{-1}1t_u^{-1}$ (T) | 0 |
| Total | 656.1 | | 673.0 |

*(S) and (T) denote singlet and triplet decay channels, respectively.

## CONCLUDING REMARKS

We reported an *ab initio* study of K-edge Auger decay in the zinc atom and the hexaaqua-zinc (II) complex. In total, 196 decay channels were studied for the atom and 1849 decay channels for the complex. Partial decay widths were computed with CBF-CCSD and CBF-MP2, while we used SFX2C-1e/EOM-CCSD for the computation of the Auger electron energies.

The K-edge Auger spectrum of zinc consists of three well-separated parts corresponding to K-LL, K-LM+K-LN, and K-MM+K-MN+K-NN transitions. Our computed ionization energies and branching ratios are in excellent agreement with previous experimental and theoretical results. The theoretical spectra for the three parts are in overall good agreement with the experimental spectra as well. The 16 K-LL transitions account for ca. 80% of the overall decay width, whereas the much more numerous K-LM and K-MM transitions contribute only 19% and 1% of intensity, respectively.

Our results demonstrate the general suitability of complex-scaled coupled-cluster methods for describing Auger decay in heavier elements beyond the second period of the periodic table but it demonstrates a significant deficiency at the same time: our theoretical approach considers relativistic effects only for the peak positions but not for the peak intensities and, therefore, does not capture the spin-orbit splitting of triplet decay channels, which leads to significant discrepancies with the experimental spectra. We also found that the presence of a water shell broadens the K-LL decay channels of zinc somewhat, whereas the water molecules do not contribute directly to the Auger spectrum. The K-edge Auger spectrum of the hexaaqua-zinc (II) complex thus is very similar to that of the bare zinc atom.

In sum, our work represents an important step towards using *ab initio* complex-variable methods for the study of Auger decay in compounds containing heavier elements that are relevant for Auger radiotherapy. The development of suitable basis sets and the consideration of spin-orbit coupling in complex-scaled calculations appear to be two necessary further steps towards this goal.

## ASSOCIATED CONTENT

**Supporting Information.** The following file is available free of charge "Supporting Information.pdf":

Basis set aug-cc-pCVTZ(5sp)+4(spd)2f in Q-Chem format used for complex-variable calculations on the core-ionized zinc atom; comparison between measured Auger spectra and Auger spectra computed using equal widths for every decay channel; partial decay half-widths computed with CBF-CCSD and CBF-MP2 for all K-edge decay channels; energies of all relevant doubly ionized states; plots of the molecular orbitals of the hexaaqua-zinc (II) complex; comparison between the half-widths of the zinc atom and the hexaaqua-zinc (II) complex.

## AUTHOR INFORMATION

**Corresponding author**

Anthuan Ferino-Pérez - *Department of Chemistry, KU Leuven, B-3001 Leuven, Belgium*; Email: anthuan.ferinoperez@kuleuven.be.

**Author**

Thomas-C. Jagau - *Department of Chemistry, KU Leuven, B-3001 Leuven, Belgium.*

**Notes**

The authors declare no competing financial interest.

## ACKNOWLEDGMENTS

The authors thank Florian Matz and Jan Philipp Drennhaus for support with the computation of partial Auger decay widths with the CBF-MP2 method. We also thank Lan Cheng and Cansu Utku for their help with running relativistic calculations with the CFOUR program package and Valentina Parravicini for useful discussions about decay processes in aqueous solutions. T.-C.J. gratefully acknowledges funding from the European Research Council (ERC) under the European Union's Horizon 2020 research and innovation program (grant agreement no. 851766) and from the KU Leuven internal funds (grant C14/22/083).

## REFERENCES

(1) Meitner, L., Über die β-Strahl-Spektra und ihren Zusammenhang mit der γ-Strahlung. *Z. Phys.* **1922,** *11* (1), 35-54.

(2) Auger, P., Sur les rayons β; secondaires produits dans un gaz par des rayons X. *C. R. Acad. Sci.* **1923,** *177*, 169-171.

(3) Agarwal, B. K., *X-ray spectroscopy: an introduction*. Springer: 2013; Vol. 15.

(4) Kassis, A. I., The Amazing World of Auger Electrons. *Int. J. Radiat Biol.* **2004,** *80* (11-12), 789-803.

(5) Brown, G. S.; Chen, M. H.; Crasemann, B.; Ice, G. E., Observation of the Auger Resonant Raman Effect. *Phys. Rev. Lett.* **1980,** *45* (24), 1937-1940.

(6) Armen, G. B.; Aksela, H.; Åberg, T.; Aksela, S., The resonant Auger effect. *J. Phys. B: At., Mol. Opt. Phys.* **2000,** *33* (2), R49-R92.

(7) Carlson, T. A.; Krause, M. O., Experimental Evidence for Double Electron Emission in an Auger Process. *Phys. Rev. Lett.* **1965,** *14* (11), 390-392.

(8) Müller, A.; Borovik, A.; Buhr, T.; Hellhund, J.; Holste, K.; Kilcoyne, A. L. D.; Klumpp, S.; Martins, M.; Ricz, S.; Viefhaus, J.; Schippers, S., Observation of a Four-Electron Auger Process in Near-*K*-Edge Photoionization of Singly Charged Carbon Ions. *Phys. Rev. Lett.* **2015,** *114* (1), 013002.

(9) Weightman, P., X-ray-excited Auger and photoelectron spectroscopy. In *Electronic Properties of Surfaces*, 2018; pp 135-195.

(10) Li, Z.; Becker, U., Chemical state effects on the Auger transitions in Cr, Fe, and Cu compounds. *J. Electron. Spectrosc. Relat. Phenom.* **2019,** *237*, 146893.

(11) Gunawardane, R. P.; Arumainayagam, C. R., Auger electron spectroscopy. In *Handbook of Applied Solid State Spectroscopy*, 2006; pp 451-483.

(12) Hofmann, S., *Auger and X-ray photoelectron spectroscopy in materials science: a user-oriented guide*. Springer Science & Business Media: 2012; Vol. 49.

(13) Raman, S. N.; Paul, D. F.; Hammond, J. S.; Bomben, K. D., Auger Electron Spectroscopy and Its Application to Nanotechnology. *Microscopy Today* **2011,** *19* (2), 12-15.

(14) Brabazon, D.; Raffer, A., Chapter 3 - Advanced characterization techniques for nanostructures. In *Emerging Nanotechnologies for Manufacturing (Second Edition)*, Ahmed, W.; Jackson, M. J., Eds. William Andrew Publishing: Boston, 2015; pp 53-85.

(15) Unger, W. E. S.; Wirth, T.; Hodoroaba, V.-D., Chapter 4.3.2 - Auger electron spectroscopy. In *Characterization of Nanoparticles*, Hodoroaba, V.-D.; Unger, W. E. S.; Shard, A. G., Eds. Elsevier: 2020; pp 373-395.

(16) Rye, R. R.; Houston, J. E., Molecular Auger spectroscopy. *Acc. Chem. Res.* **1984,** *17* (1), 41-47.

(17) Marchenko, T.; Inhester, L.; Goldsztejn, G.; Travnikova, O.; Journel, L.; Guillemin, R.; Ismail, I.; Koulentianos, D.; Céolin, D.; Püttner, R.; Piancastelli, M. N.; Simon, M., Ultrafast nuclear dynamics in the doubly-core-ionized water molecule observed via Auger spectroscopy. *Phys. Rev. A* **2018,** *98* (6), 063403.

(18) Khokhlova, M. A.; Cooper, B.; Ueda, K.; Prince, K. C.; Kolorenč, P.; Ivanov, M. Y.; Averbukh, V., Molecular Auger Interferometry. *Phys. Rev. Lett.* **2019,** *122* (23), 233001.

(19) Plekan, O.; Sa'adeh, H.; Ciavardini, A.; Callegari, C.; Cautero, G.; Dri, C.; Di Fraia, M.; Prince, K. C.; Richter, R.; Sergo, R.; Stebel, L.; Devetta, M.; Faccialà, D.; Vozzi, C.; Avaldi, L.; Bolognesi, P.; Castrovilli, M. C.; Catone, D.; Coreno, M.; Zuccaro, F.; Bernes, E.; Fronzoni, G.; Toffoli, D.; Ponzi, A., Experimental and Theoretical Photoemission Study of Indole and Its Derivatives in the Gas Phase. *J. Phys. Chem. A* **2020,** *124* (20), 4115-4127.

(20) Kassis, A. I.; Adelstein, S. J., Radiobiologic Principles in Radionuclide Therapy. *J. Nucl. Med.* **2005,** *46* (1 Suppl), 4S-12S.

(21) Buchegger, F.; Perillo-Adamer, F.; Dupertuis, Y. M.; Bischof Delaloye, A., Auger radiation targeted into DNA: a therapy perspective. *Eur. J. Nucl. Med. Mol. Imaging* **2006,** *33* (11), 1352-1363.

(22) Vallis, K. A.; Reilly, R. M.; Scollard, D.; Merante, P.; Brade, A.; Velauthapillai, S.; Caldwell, C.; Chan, I.; Freeman, M.; Lockwood, G.; Miller, N. A.; Cornelissen, B.; Petronis, J.; Sabate, K., Phase I trial to evaluate the tumor and normal tissue uptake, radiation dosimetry and safety of (111)In-DTPA-human epidermal growth factor in patients with metastatic EGFR-positive breast cancer. *Am. J. Nucl. Med. Mol. Imaging* **2014,** *4* (2), 181-192.

(23) Li, L.; Quang, T. S.; Gracely, E. J.; Kim, J. H.; Emrich, J. G.; Yaeger, T. E.; Jenrette, J. M.; Cohen, S. C.; Black, P.; Brady, L. W., A phase II study of anti-epidermal growth factor receptor radioimmunotherapy in the treatment of glioblastoma multiforme. *J. Neurosurg.* **2010,** *113* (2), 192-198.

(24) Ku, A.; Facca, V. J.; Cai, Z.; Reilly, R. M., Auger electrons for cancer therapy – a review. *EJNMMI Radiopharm. Chem.* **2019,** *4* (1), 27.

(25) Pirovano, G.; Wilson, T. C.; Reiner, T., Auger: The future of precision medicine. *Nucl. Med. Biol.* **2021,** *96-97*, 50-53.

(26) Howell, R. W., Radiation spectra for Auger-electron emitting radionuclides: Report No. 2 of AAPM Nuclear Medicine Task Group No. 6. *Med. Phys.* **1992,** *19* (6), 1371-1383.

(27) Moiseyev, N., *Non-Hermitian quantum mechanics*. Cambridge University Press: 2011.

(28) Jagau, T.-C., Theory of electronic resonances: fundamental aspects and recent advances. *Chem. Commun.* **2022,** *58* (34), 5205-5224.

(29) Liu, W., Essentials of relativistic quantum chemistry. *J. Chem. Phys.* **2020,** *152* (18).

(30) Chen, M. H.; Crasemann, B.; Mark, H., Relativistic *K-LL* Auger spectra in the intermediate-coupling scheme with configuration interaction. *Phys. Rev. A* **1980,** *21* (2), 442-448.

(31) Chen, M. H.; Crasemann, B.; Mark, H., Relativistic *K*-shell Auger rates, level widths, and fluorescence yields. *Phys. Rev. A* **1980,** *21* (2), 436-441.

(32) Safronova, U. I.; Johnson, W. R.; Albritton, J. R., Auger rates for Ni-, Cu-, and Zn-like ions. *At. Data Nucl. Data Tables* **2001,** *77* (2), 215-275.

(33) Martins, L.; Amaro, P.; Pessanha, S.; Guerra, M.; Machado, J.; Carvalho, M. L.; Santos, J. P., Multiconfiguration Dirac–Fock calculations of Zn K-shell radiative and nonradiative transitions. *X-Ray Spectrom.* **2020,** *49* (1), 192-199.

(34) Melia, H. A.; Dean, J. W.; Nguyen, T. V. B.; Chantler, C. T., Cu K$\alpha_{3,4}$ satellite spectrum with ab initio Auger-rate calculations. *Phys. Rev. A* **2023,** *107* (1), 012809.

(35) Fano, U., Effects of Configuration Interaction on Intensities and Phase Shifts. *Phys. Rev.* **1961,** *124* (6), 1866-1878.

(36) Feshbach, H., A unified theory of nuclear reactions. II. *Annals of Physics* **1962,** *19* (2), 287-313.

(37) Matz, F.; Jagau, T.-C., Molecular Auger decay rates from complex-variable coupled-cluster theory. *J. Chem. Phys.* **2022,** *156* (11), 114117.

(38) Matz, F.; Jagau, T.-C., Channel-specific core-valence projectors for determining partial Auger decay widths. *Mol. Phys.* **2022**, e2105270.

(39) Jayadev, N. K.; Ferino-Pérez, A.; Matz, F.; Krylov, A. I.; Jagau, T.-C., The Auger spectrum of benzene. *J. Chem. Phys.* **2023,** *158* (6), 064109.

(40) Matz, F.; Nijssen, J.; Jagau, T.-C., Ab Initio Investigation of the Auger Spectra of Methane, Ethane, Ethylene, and Acetylene. *J. Phys. Chem. A* **2023,** *127* (30), 6147-6158.

(41) Aguilar, J.; Combes, J. M., A class of analytic perturbations for one-body Schrödinger Hamiltonians. *Commun. Math. Phys.* **1971,** *22* (4), 269-279.

(42) Balslev, E.; Combes, J.-M., Spectral properties of many-body Schrödinger operators with dilatation-analytic interactions. *Commun. Math. Phys.* **1971,** *22*, 280-294.

(43) McCurdy, C. W.; Rescigno, T. N., Extension of the Method of Complex Basis Functions to Molecular Resonances. *Phys. Rev. Lett.* **1978,** *41* (20), 1364-1368.

(44) Rescigno, T. N.; Orel, A. E.; McCurdy, C. W., Application of complex coordinate SCF techniques to a molecular shape resonance: The $^2\Pi_g$ state of $N_2^-$. *J. Chem. Phys.* **1980,** *73* (12), 6347-6348.

(45) White, A. F.; Head-Gordon, M.; McCurdy, C. W., Complex basis functions revisited: Implementation with applications to carbon tetrafluoride and aromatic N-containing heterocycles within the static-exchange approximation. *J. Chem. Phys.* **2015,** *142* (5), 054103.

(46) Roos, B. O.; Lindh, R.; Malmqvist, P.-Å.; Veryazov, V.; Widmark, P.-O., New Relativistic ANO Basis Sets for Transition Metal Atoms. *J. Phys. Chem. A* **2005,** *109* (29), 6575-6579.

(47) Jana, C.; Ohanessian, G.; Clavaguéra, C., Theoretical insight into the coordination number of hydrated $Zn^{2+}$ from gas phase to solution. *Theor. Chem. Acc.* **2016,** *135* (5), 141.

(48) Marcus, Y., Ionic radii in aqueous solutions. *Chem. Rev.* **1988,** *88* (8), 1475-1498.

(49) Weigend, F.; Häser, M.; Patzelt, H.; Ahlrichs, R., RI-MP2: optimized auxiliary basis sets and demonstration of efficiency. *Chem. Phys. Lett.* **1998,** *294* (1), 143-152.

(50) Emrich, K., An extension of the coupled cluster formalism to excited states (I). *Nucl. Phys. A* **1981,** *351* (3), 379-396.

(51) Nooijen, M.; Snijders, J. G., Coupled cluster Green's function method: Working equations and applications. *Int. J. Quantum Chem* **1993,** *48* (1), 15-48.

(52) Stanton, J. F.; Bartlett, R. J., The equation of motion coupled-cluster method. A systematic biorthogonal approach to molecular excitation energies, transition probabilities, and excited state properties. *J. Chem. Phys.* **1993,** *98* (9), 7029-7039.

(53) Stanton, J. F.; Gauss, J., Analytic energy derivatives for ionized states described by the equation-of-motion coupled cluster method. *J. Chem. Phys.* **1994,** *101* (10), 8938-8944.

(54) Shavitt, I.; Bartlett, R. J., *Many-body methods in chemistry and physics: MBPT and coupled-cluster theory*. Cambridge University Press: 2009.

(55) Nooijen, M.; Bartlett, R. J., Similarity transformed equation-of-motion coupled-cluster theory: Details, examples, and comparisons. *J. Chem. Phys.* **1997,** *107* (17), 6812-6830.

(56) Sattelmeyer, K. W.; Schaefer III, H. F.; Stanton, J. F., Use of 2h and 3h−p-like coupled-cluster Tamm–Dancoff approaches for the equilibrium properties of ozone. *Chem. Phys. Lett.* **2003,** *378* (1), 42-46.

(57) Dyall, K. G., Interfacing relativistic and nonrelativistic methods. IV. One- and two-electron scalar approximations. *J. Chem. Phys.* **2001,** *115* (20), 9136-9143.

(58) Cheng, L.; Gauss, J., Perturbative treatment of spin-orbit coupling within spin-free exact two-component theory. *J. Chem. Phys.* **2014,** *141* (16), 164107.

(59) Cheng, L.; Wang, F.; Stanton, J. F.; Gauss, J., Perturbative treatment of spin-orbit-coupling within spin-free exact two-component theory using equation-of-motion coupled-cluster methods. *J. Chem. Phys.* **2018,** *148* (4), 044108.

(60) Stanton, J. F.; Gauss, J., A simple scheme for the direct calculation of ionization potentials with coupled-cluster theory that exploits established excitation energy methods. *J. Chem. Phys.* **1999,** *111* (19), 8785-8788.

(61) Matthews, D. A.; Cheng, L.; Harding, M. E.; Lipparini, F.; Stopkowicz, S.; Jagau, T.-C.; Szalay, P. G.; Gauss, J.; Stanton, J. F., Coupled-cluster techniques for computational chemistry: The CFOUR program package. *J. Chem. Phys.* **2020,** *152* (21), 214108.

(62) Epifanovsky, E.; Gilbert, A. T. B.; Feng, X.; Lee, J.; Mao, Y.; Mardirossian, N.; Pokhilko, P.; White, A. F.; Coons, M. P.; Dempwolff, A. L.; Gan, Z.; Hait, D.; Horn, P. R.; Jacobson, L. D.; Kaliman, I.; Kussmann, J.; Lange, A. W.; Lao, K. U.; Levine, D. S.; Liu, J.; McKenzie, S. C.; Morrison, A. F.; Nanda, K. D.; Plasser, F.; Rehn, D. R.; Vidal, M. L.; You, Z.-Q.; Zhu, Y.; Alam, B.; Albrecht, B. J.; Aldossary, A.; Alguire, E.; Andersen, J. H.; Athavale, V.; Barton, D.; Begam, K.; Behn, A.; Bellonzi, N.; Bernard, Y. A.; Berquist, E. J.; Burton, H. G. A.; Carreras, A.; Carter-Fenk, K.; Chakraborty, R.; Chien, A. D.; Closser, K. D.; Cofer-Shabica, V.; Dasgupta, S.; Wergifosse, M. d.; Deng, J.; Diedenhofen, M.; Do, H.; Ehlert, S.; Fang, P.-T.; Fatehi, S.; Feng, Q.; Friedhoff, T.; Gayvert, J.; Ge, Q.; Gidofalvi, G.; Goldey, M.; Gomes, J.; González-Espinoza, C. E.; Gulania, S.; Gunina, A. O.; Hanson-Heine, M. W. D.; Harbach, P. H. P.; Hauser, A.; Herbst, M. F.; Vera, M. H.; Hodecker, M.; Holden, Z. C.; Houck, S.; Huang, X.; Hui, K.; Huynh, B. C.; Ivanov, M.; Jász, Á.; Ji, H.; Jiang, H.; Kaduk, B.; Kähler, S.; Khistyaev, K.; Kim, J.; Kis, G.; Klunzinger, P.; Koczor-Benda, Z.; Koh, J. H.; Kosenkov, D.; Koulias, L.; Kowalczyk, T.; Krauter, C. M.; Kue, K.; Kunitsa, A.; Kus, T.; Ladjánszki, I.; Landau, A.; Lawler, K. V.; Lefrancois, D.; Lehtola, S.; Li, R. R.; Li, Y.-P.; Liang, J.; Liebenthal, M.; Lin, H.-H.; Lin, Y.-S.; Liu, F.; Liu, K.-Y.; Loipersberger, M.; Luenser, A.; Manjanath, A.; Manohar, P.; Mansoor, E.; Manzer, S. F.; Mao, S.-P.; Marenich, A. V.; Markovich, T.; Mason, S.; Maurer, S. A.; McLaughlin, P. F.; Menger, M. F. S. J.; Mewes, J.-M.; Mewes, S. A.; Morgante, P.; Mullinax, J. W.; Oosterbaan, K. J.; Paran, G.; Paul, A. C.; Paul, S. K.; Pavošević, F.; Pei, Z.; Prager, S.; Proynov, E. I.; Rák, Á.; Ramos-Cordoba, E.; Rana, B.; Rask, A. E.; Rettig, A.; Richard, R. M.; Rob, F.; Rossomme, E.; Scheele, T.;

Scheurer, M.; Schneider, M.; Sergueev, N.; Sharada, S. M.; Skomorowski, W.; Small, D. W.; Stein, C. J.; Su, Y.-C.; Sundstrom, E. J.; Tao, Z.; Thirman, J.; Tornai, G. J.; Tsuchimochi, T.; Tubman, N. M.; Veccham, S. P.; Vydrov, O.; Wenzel, J.; Witte, J.; Yamada, A.; Yao, K.; Yeganeh, S.; Yost, S. R.; Zech, A.; Zhang, I. Y.; Zhang, X.; Zhang, Y.; Zuev, D.; Aspuru-Guzik, A.; Bell, A. T.; Besley, N. A.; Bravaya, K. B.; Brooks, B. R.; Casanova, D.; Chai, J.-D.; Coriani, S.; Cramer, C. J.; Cserey, G.; DePrinceIII, A. E.; DiStasioJr., R. A.; Dreuw, A.; Dunietz, B. D.; Furlani, T. R.; GoddardIII, W. A.; Hammes-Schiffer, S.; Head-Gordon, T.; Hehre, W. J.; Hsu, C.-P.; Jagau, T.-C.; Jung, Y.; Klamt, A.; Kong, J.; Lambrecht, D. S.; Liang, W.; Mayhall, N. J.; McCurdy, C. W.; Neaton, J. B.; Ochsenfeld, C.; Parkhill, J. A.; Peverati, R.; Rassolov, V. A.; Shao, Y.; Slipchenko, L. V.; Stauch, T.; Steele, R. P.; Subotnik, J. E.; Thom, A. J. W.; Tkatchenko, A.; Truhlar, D. G.; Voorhis, T. V.; Wesolowski, T. A.; Whaley, K. B.; WoodcockIII, H. L.; Zimmerman, P. M.; Faraji, S.; Gill, P. M. W.; Head-Gordon, M.; Herbert, J. M.; Krylov, A. I., Software for the frontiers of quantum chemistry: An overview of developments in the Q-Chem 5 package. *J. Chem. Phys.* **2021,** *155* (8), 084801.

(63) Bravaya, K. B.; Zuev, D.; Epifanovsky, E.; Krylov, A. I., Complex-scaled equation-of-motion coupled-cluster method with single and double substitutions for autoionizing excited states: Theory, implementation, and examples. *J. Chem. Phys.* **2013,** *138* (12), 124106.

(64) Zuev, D.; Jagau, T.-C.; Bravaya, K. B.; Epifanovsky, E.; Shao, Y.; Sundstrom, E.; Head-Gordon, M.; Krylov, A. I., Complex absorbing potentials within EOM-CC family of methods: Theory, implementation, and benchmarks. *J. Chem. Phys.* **2014,** *141* (2), 024102.

(65) White, A. F.; Epifanovsky, E.; McCurdy, C. W.; Head-Gordon, M., Second order Møller-Plesset and coupled cluster singles and doubles methods with complex basis functions for resonances in electron-molecule scattering. *J. Chem. Phys.* **2017,** *146* (23), 234107.

(66) Moiseyev, N.; Certain, P. R.; Weinhold, F., Resonance properties of complex-rotated hamiltonians. *Mol. Phys.* **1978,** *36* (6), 1613-1630.

(67) Jagau, T.-C., Coupled-cluster treatment of molecular strong-field ionization. *J. Chem. Phys.* **2018,** *148* (20), 204102.

(68) Dunning, T. H., Jr., Gaussian basis sets for use in correlated molecular calculations. I. The atoms boron through neon and hydrogen. *J. Chem. Phys.* **1989,** *90* (2), 1007-1023.

(69) Kendall, R. A.; Dunning, T. H., Jr.; Harrison, R. J., Electron affinities of the first-row atoms revisited. Systematic basis sets and wave functions. *J. Chem. Phys.* **1992,** *96* (9), 6796-6806.

(70) Woon, D. E.; Dunning, T. H., Jr., Gaussian basis sets for use in correlated molecular calculations. V. Core-valence basis sets for boron through neon. *J. Chem. Phys.* **1995,** *103* (11), 4572-4585.

(71) Balabanov, N. B.; Peterson, K. A., Systematically convergent basis sets for transition metals. I. All-electron correlation consistent basis sets for the 3d elements Sc–Zn. *J. Chem. Phys.* **2005,** *123* (6), 064107.

(72) Peterson, K. A.; Dunning, T. H., Jr., Accurate correlation consistent basis sets for molecular core–valence correlation effects: The second row atoms Al–Ar, and the first row atoms B–Ne revisited. *J. Chem. Phys.* **2002,** *117* (23), 10548-10560.

(73) Kahk, J. M.; Lischner, J., Accurate absolute core-electron binding energies of molecules, solids, and surfaces from first-principles calculations. *Physical Review Materials* **2019,** *3* (10), 100801.

(74) Freedman, M. S.; Porter, F. T.; Wagner, F., Internal Conversion, Multipole Mixing, and Auger Spectrum in $Zn^{67}$ from $Ga^{67}$ Decay. *Phys. Rev.* **1966,** *151* (3), 886-898.

(75) Shirley, D. A.; Martin, R. L.; Kowalczyk, S. P.; McFeely, F. R.; Ley, L., Core-electron binding energies of the first thirty elements. *Phys. Rev. B* **1977,** *15* (2), 544-552.

(76) Kovalík, A.; Lubashevsky, A. V.; Inoyatov, A.; Filosofov, D. V.; Korolev, N. A.; Gorozhankin, V. M.; Vylov, T.; Štekl, I., A detailed experimental investigation of the low energy electron spectrum generated in the EC-decay of $^{67}$Ga. *J. Electron. Spectrosc. Relat. Phenom.* **2004,** *134* (1), 67-79.

(77) Inoyatov, A. K.; Perevoshchikov, L. L.; Gorozhankin, V. M.; Kovalík, A.; Radchenko, V. I.; Filosofov, D. V., Searching for influence of the atomic structure effect on the KLL and LMM Auger transition energies of Zn (Z=30) and Gd (Z=64). *J. Electron. Spectrosc. Relat. Phenom.* **2011,** *184* (8), 457-462.

(78) Babenkov, M.; Bobykin, B.; Petukhov, V. *Dependence of the absolute probabilities of KLL Auger transitions on the atomic number, 30 less than or equal to Z less than or equal to 94. II*; Inst. of Nuclear Physics, Alma-Ata, USSR: 1972.

(79) Tarantelli, F.; Sgamellotti, A.; Cederbaum, L. S.; Schirmer, J., Theoretical investigation of many dicationic states and the Auger spectrum of benzene. *J. Chem. Phys.* **1987,** *86* (4), 2201-2206.

(80) Tarantelli, F.; Sgamellotti, A.; Cederbaum, L. S., Many dicationic states and two-hole population analysis as a bridge to Auger spectra: Strong localization phenomena in BF3. *J. Chem. Phys.* **1991,** *94* (1), 523-532.

(81) Bruneau, J., MCDF calculation of argon Auger process. *J. Phys. B* **1983,** *16* (22), 4135.

(82) Karim, K. R.; Crasemann, B., $L_1$-$L_{23}M_1$ Coster-Kronig spectrum of argon in intermediate coupling. *Phys. Rev. A* **1984,** *30* (2), 1107-1108.

(83) Karim, K. R.; Crasemann, B., Continuum interaction in low-energy radiationless transitions. *Phys. Rev. A* **1985,** *31* (2), 709-713.

(84) Gopakumar, G.; Unger, I.; Slavíček, P.; Hergenhahn, U.; Öhrwall, G.; Malerz, S.; Céolin, D.; Trinter, F.; Winter, B.; Wilkinson, I.; Caleman, C.; Muchová, E.; Björneholm, O., Radiation damage by extensive local water ionization from two-step electron-transfer-mediated decay of solvated ions. *Nat. Chem.* **2023,** *15* (10), 1408-1414.

(85) Stumpf, V.; Gokhberg, K.; Cederbaum, L. S., The role of metal ions in X-ray-induced photochemistry. *Nat. Chem.* **2016,** *8* (3), 237-241.

**TOC abstract graphic**

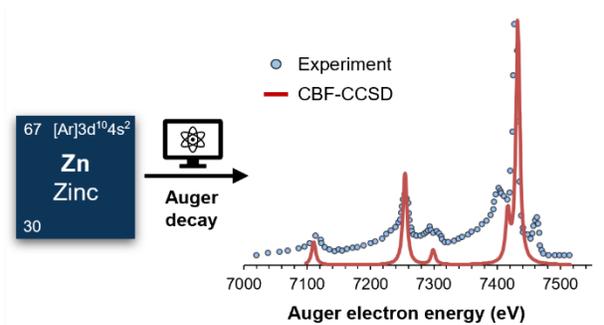

For Table of Contents Only